\documentstyle[preprint,version2,aps]{revtex} 
 \begin{document}
 \tightenlines
 \draft

\begin{title}
 Double gap and solitonic excitations in the spin-Peierls chain  CuGeO$_3$
 \end{title}

\author{M. A\"\i n$^{1*}$, J.E. Lorenzo$^2$, L. P.~Regnault$^3$,
 G.~Dhalenne$^4$, A.~Revcolevschi$^4$, B. Hennion$^1$ and Th.~Jolicoeur$^5$.}

\begin{instit}
$^1$Laboratoire L\'eon Brillouin (CEA-CNRS) CE-Saclay. 91191 Gif/Yvette C\'edex
FRANCE. 
\end{instit}

\begin{instit}
$^2$European Synchroton Radiation Facility, 38043 Grenoble FRANCE. 
\end{instit}

\begin{instit}
$^3$Laboratoire de Magn\'etisme et de Diffraction Neutronique, CENG. 38054
 Grenoble C\'edex 9 FRANCE. 
\end{instit}

\begin{instit}
$^4$Laboratoire de Chimie des Solides (Bat 414) - UA CNRS 446 Universit\'e de
Paris-Sud. 91405 Orsay C\'edex FRANCE. 
\end{instit}

\begin{instit}
$^5$Service de Physique Th\'eorique, CE-Saclay. 91191 Gif/Yvette C\'edex FRANCE.
\end{instit}

%%%%%%%%%%%%%%%%%%%%%%%%%%%%%%%%%%%%%%%%%%%%%%%%%%%%%%%%%%%%%

\begin{abstract}
 We have studied magnetic excitations in the dimerized spin-Peierls phase 
of CuGeO$_3$, by high resolution inelastic neutron scattering. We measured the
well-defined spin triplet dispersive mode which is gapped throughout the whole
Brillouin zone. We also observed that this mode is separated by an unexpected
second gap of $\approx$ 2 meV from a continuum of magnetic excitations extending
to higher energy.  The first gap (or 'triplet gap') and its associated
dispersive mode is due to the breaking of a singlet dimer into a delocalized
triplet.  We propose that the second gap (or 'solitonic gap') and the continuum
correspond to dissociation of that triplet into two unbound spin-1/2 solitons
that are separated by a dimerized region of arbitrary length.\end{abstract}

%%%%%%%%%%%%%%%%%%%%%%%%%%%%%%%%%%%%%%%%%%%%%%%%%%%%%%%%%%%%%


\narrowtext

The recent observation by Hase\cite{hase} of a characteristic magnetic
susceptibility in CuGeO$_3$, dropping abruptly to zero below T$_{SP}$=14.3 K,
clearly suggested that it was a new one-dimensional spin-Peierls compound. This
was confirmed by X-Rays photographs\cite{pouget,hiro} that revealed, below
T$_{SP}$, superlattice peaks indexing according to the propagation vector
k$_{SP}$=(0.5,0,0.5). These two experimental evidences indicate : 1- That a gap
has opened over a non magnetic singlet ground state as demonstrated by neutron
studies\cite{nish,reg} and 2- That the crystal was undergoing a magnetoelastic
distortion where copper ions dimerize with their left or right nearest neighbor
along the chains. As a result, the initially uniform exchange coupling becomes
staggered.

Single crystals of CuGeO$_3$ are grown in an image furnace by the traveling
floating zone method. They belong to the orthorhombic space group Pbmm. Magnetic
chains of Cu$^{++}$, S=1/2 ions are parallel to the c axis. The spin-Peierls gap
is observable at k$_{AF}$=(0,0,0.5) or equivalent points, but there are no
magnetic Bragg peaks. Dispersion curves of magnetic excitations along the three
principal directions a,b and c are of simple sinusoidal shape\cite{nish,reg};
estimates for intrachain nearest neighbors exchange (NNE) along c gave
$J_1\approx$ 120 K, as derived from fits to classical magnon theory in
ref\cite{nish} or to Bonner-Bl\"ote relation in ref\cite{reg}. Fits to classical
magnon theory\cite{nish,reg}, indicate that NNE between chains are only one and
two orders of magnitude smaller along the b and a directions
respectively.

In this letter we present experimental evidence that there are in fact two gaps
in this system and not only one as predicted by the classical
approach\cite{bonner,Cross}. Although this observation could fit in the
framework of Cross-Fisher theory\cite{Cross}, we shall use a solitonic approach
to elaborate an explanation which accounts for two gaps. A dimerized system has
an obvious excitation which consists of breaking a dimer bond into a triplet at
a cost of a certain magnetoelastic energy. The triplet will be delocalized along
the chain generating eigenstates of definite momentum.  However there is another
possible excitation in this system because the triplet can absorb a second
amount of energy corresponding to a second gap and thus dissociate into two
S=1/2 traveling solitons that generate the continuum. This has some analogies to
the well known two-spinon continuum of the uniform (undimerized) Heisenberg
S=1/2 AF chain (HAFC) that has been investigated extensively in KCuF$_3$ (see
figure 6 of ref\cite{tennant}). Above $T_{SP}$ the analogy is even more complete
as we shall see on figure~\ref{corbose}. On the other hand it has been suggested
that competing next-nearest-neighbor exchange (NNNE) was the driving mechanism
in the dimerization process of CuGeO$_3$ instead of magnetoelastic coupling as
generally admitted. In support of this idea, were satisfactory
fits\cite{CCE,Riera,zang} of the magnetic susceptibility of CuGeO$_3$ above
T$_{SP}$. Since this susceptibility was not well reproduced by the Bonner-Fisher
curve\cite{BF}, which is appropriate to the isolated S=1/2 AF NNE chain,
incorporation of NNNE gave better results. Yet in the final part of section II
of our discussion we express some reservation with respect to this
interpretation based on competing NNNE.

Inelastic neutron scattering measurements have been performed on two triple-axis
spectrometers : 1- 4F1 (Orph\'ee reactor, LLB Saclay) has an incident beam,
fixed in direction, extracted by a pair of graphite monochromators (the second
one being vertically focussing). It was operated at constant k$_f$=1.55
${\AA}^{-1}$ (5.01 meV) with a horizontally focussing graphite analyser and a
berylium filter to cut out higher-order components of the diffracted beam.  2-
IN14 (HFR, ILL Grenoble) has one vertically focussing graphite monochromator and
a horizontally focussing graphite analyzer, the rest of the set up was similar
on both spectrometers. The resolution on both apparatus was very close to 0.2
meV (FWHM) as deduced from the incoherent peak at zero energy transfer. In both
experiments the same single crystal (nearly 1 cm$^3$) was oriented with the b
and c crystallographic axes in the scattering plane. Two series of inelastic
scans were recorded for neutron energy transfers ranging from -0.3 meV to 11.5
meV. All scans are corrected for $\lambda /2$ contamination in the incident
beam\cite{fak}.

Figure~(\ref{ccont}) shows three energy scans at T=2.6 K. They correspond to
excitations near the zone boundary along the c$*$ direction parallel to the
chains. Five elements are visible on the scan at Q=(0,1,0.5) : 1- The
zero-energy incoherent peak showing the spectrometer resolution. 2- A first gap,
called hereafter the 'triplet gap' with a value of $\Delta = 2$ meV at
$Q=(0,1,0.5)$.  3- A well-defined magnon-like mode first observed by Nishi {\it
et al.}.\onlinecite{nish}.  This mode is in fact a spin-triplet mode as shown by
measurements in a magnetic field\cite{reg}. Its asymmetric shape is due to
convolution of instrumental resolution and steep curvature of the dispersion
curve in the vicinity of Q=(0,1,0.5).  4- The intensity between the middle peak
(or triplet mode) and the plateau, falls to the background level. This is
clearly a new gap in energy that we call hereafter the 'solitonic gap'. At
Q=(0,1,0.5), this 'solitonic gap' is close to 2 meV. Defining the background in
this experiment, is an issue that will be addressed when presenting
figure~(\ref{corbose}). 5- Finally, brought out by the 'solitonic gap' we find
some intensity (low but clearly present) which constitute the expected continuum
that extends at least up to the maximum energy transfer of our study, i.e. 11.5
meV. Scans for Q=(0,1,0.48) and Q=(0,1,0.46) display the same structure as that
for Q=(0,1,0.5) with the exception of the incoherent peak which is not shown. We
recall that neutron\cite{reg,arai} and Ramann\cite{loosd} scattering have
already given clear evidence for the existence of such a continuum.

Figure~(\ref{bcont}) shows a series of six energy scans regularly spaced along
b$^*$ between Q=(0,1,0.5) and Q=(0,2,0.5) at T=1.7 K. No incoherent peak here,
only the peak of the dispersive mode followed again by the 'solitonic gap' and
the continuum are visible in this series of scans. Note that owing to coupling
between chains, as already mentionned,  there is dispersion along Q$_b$ and the positions of the peaks
are not constant in energy as they would be for a pure one-dimensional system.

Figure~(\ref{corbose}) provides a more detailed picture of the energy scans at
Q=(0,1,0.5). It shows that when the temperature is raised to T=29 K the peak of
the dispersive mode drops and widens, filling in the double gap region and
merging with the continuum indicating that the two gaps have collapsed and that
we have recovered the continuum of the uniform HAFC. When we reach T=150 K, the
magnon-like mode and the continuum have totally disappeared, we are then in the
truly paramagnetic region; note that the intensity falls to the level of what
was measured at T=1.7 K in either gap, this level is considered as the
background of our experiment. In the insert we subtracted the scan at 150 K from the scan at 1.7
K, both were corrected for Bose factor after subtraction of a background of 15
counts on each; what remains is the dispersive mode, the 'solitonic gap' and the
continuum. A phonon is distinguishable near 11 meV in the 150 K data.

The fact that the double gap has been overlooked in previous
experiments\cite{nish,reg,arai} is due to poorer resolution of the instruments
used before. In these former experiments the high energy tail of the
uncompletely resolved dispersive mode precluded observation of the 'solitonic
gap' by causing a smooth crossover to the high energy continuum.  In the present
experiment high resolution was obtained through the use of a small k$_f$=1.55
${\AA}$.

It has been impossible to detect acoustic phonon branches around k$_{SP}$, and
moreover, preliminary polarized neutron measurements at Q=(0,1,0.5) and
Q=(0.5,5,0.5) indicate that all of the intensity in the peak of the dispersive
mode is magnetic, as is the major part if not all, of the intensity in the
continuum. The absence of an inelastic nuclear contribution is consistent with
the fact that in the dimerized phase of CuGeO$_3$, intensities of nuclear
superlattice peaks at \{k$_{SP}$\} are very weak and displacements of atoms
extremely tiny. Nonetheless superlattice nuclear peaks are sensitive to a
magnetic field as proved by the commensurate-incommensurate transition that
occurs at 12.5 T\cite{kiryu}. All this suggests that there could exist a
magnetoelastic coupling through spin-charge hybridization\cite{braden}.

There are three models that are related to our observations in CuGeO$_3$ : I -
The AF chain with NNE $J_1$ and NNNE $J_2$. II - The AF chain with NNE only but
with imposed staggering of the exchange. III - The AF chain coupled to a phonon
field $u(x)$\cite{Cross}. Although all three models predict a gap, only model
III predicts a double gap which is consistent with our experimental results
on the excitation spectrum of CuGeO$_3$.  An analysis of these models has been
given by Haldane\cite{hal}.

I - In the case of NNE $J_1$ and NNNE $J_2$, the Hamiltonian of the chain has
full translational invariance. If $J_2$ is smaller than a critical coupling
$(J_2 /J_1 < 0.2412(1))$, the ground state is a spin liquid but if $J_2$ is larger, the ground state is dimerized and twice
degenerate. Translation invariance by one lattice spacing is spontaneously
broken. We will refer to this situation as "spontaneous dimers". The effective
long-wavelength, low-energy theory is described by a sine-Gordon
model\cite{Affleck}:

\begin{equation} H= \int\, dx \, {1\over 2}(\Pi^2 + (\nabla
\phi )^2 ) +\alpha \cos (\beta \phi ).  \label{SG}
\end{equation}
In this equation, $\Pi$ is the momentum conjugate to $\phi$ which is related to
the z-component of the spin at position x by : $S^z (x) = -\nabla\phi
(x)/\sqrt{2\pi} + {\rm C}(-)^x \cos (\sqrt{2\pi}\phi (x))$. When 
$J_2\neq 0$ , the value of the sine-Gordon coupling is $\beta
=2\sqrt{2\pi}$. If $J_2$ is larger than the critical value, one is in the
massive (or gapped) phase of the theory of Eq.~(\ref{SG}). The $\beta
=2\sqrt{2\pi}$ sine-Gordon theory has no bound states\cite{Dashen} and the
elementary excitations are kinks that correspond to a $\pm 2\pi$ variation of
the argument of the cosine term in Eq.~(\ref{SG}) over a localized region of
space. These solitons therefore have spin S=1/2. The physical picture is
simple : an excitation means that a singlet in the dimerized ground state is
broken into a triplet that immediately disintegrates into two free solitons.  As
a consequence the magnetic excitations form a continuum above some threshold and
there is no well-defined mode below. This is not what we observe in our
experiments.

II - We turn now to the externally dimerized chain: there is an additional
modulated exchange $\delta \sum_n (-)^n {\vec S}_n \cdot {\vec S}_{n+1}$.  There
we have explicit doubling of the unit cell. This also
leads to a sine-Gordon model Eq.~(\ref{SG}) but now with a coupling $\beta
=\sqrt{2\pi}$.  The kinks that still correspond to a $\pm 2\pi$ variation of the
argument of the cosine in Eq.~(\ref{SG}) have now spin $S^z=\pm 1$. The
sine-Gordon theory with $\beta =\sqrt{2\pi}$ has two breather bound
states\cite{Dashen} : one which is degenerate with the kink states.  This state
completes the S=1 triplet which is expected due to the full rotational
invariance of the theory. The other bound state is a singlet and thus plays no
role in the magnetic excitation spectrum. Here the physical picture is quite
different from that of model I. The singlet bonds are pinned to the lattice by
the dimerizing potential. The elementary excitation corresponds to breaking a
singlet bond in a triplet and then this triplet will move along the chain. This
triplet state is not a domain wall and cannot disintegrate as has been seen in
numerical simulations\cite{CCE,Riera,haas}. There are continua above this
well-defined mode that are due to excitations of several triplets. Haas and
Dagotto have recently performed a study\cite{haas} of the dynamical properties
of an externally dimerized chain including a NNNE $J_2$.  They have shown that
there is a continuum starting immediately above the spin triplet mode contrary
to our finding of a 'solitonic gap' in CuGeO$_3$.

To the extent that NNNE should be visible on the shape of the dispersion curve
it becomes informative to calculate the dynamics of a colinear S=1/2 Heisenberg
AF model with $J_1 < 0$ and $J_2$, and lattice spacing c. It yields the
following dispersion relation :

\begin{equation}
h\nu (q) = 2\vert \sin {qc} \vert \sqrt{({J_1}(J_1-4J_2) + 4{J_2}^2\sin^2 {qc}}
\label{dispcrb} \end{equation} 
which is only valid for $J_2/J_1<0.25$ (Villain's criterion).  We see that the
curve $\nu (q)$ is narrowed by the last term with the $J_2$ factor, which makes
it depart markedly from a sinusoidal shape in contrast to our observations on
CuGeO$_3$. The analogy with CuGeO$_3$ is enforced by noting that the underlying
AF system in CuGeO$_3$ would have AF Bragg peaks at k$_{AF}$=(0,0,0.5) as
verified in CuGe$_{0.993}$Si$_{0.007}$O$_3$\cite{reg1} where AF and dimerization
coexist. (If $J_2 > 0$ and $J_2/J_1>0.25$ the calculation ought to be conducted
differently as appropriate for helimagnetic incommensurate AF).

III - Finally we turn to the more realistic model having an elastic displacement
field\cite{Cross,Naka}. It involves an additional coupling of the sine-Gordon
field with an elastic field, of the form $u(x)\cos(\sqrt{2\pi}\phi (x))$. The
Hamiltonian Eq.(\ref{SG}) again has full translational invariance.  This
invariance is spontaneously broken and there are thus domain walls: the
displacement $u(x)$ in a domain wall goes from $+ u_0$ on one side of the chain
to $- u_0$ on the other side, or vice-versa. The spin soliton involves only a
variation of $\pi$ of the argument of the cosine and thus has spin S=1/2, as in
the case of the spontaneously dimerized chain. This elementary soliton can be
visualized as an isolated S=1/2 copper spin that separates two regions of the
chain that are dimerized.  These solitons have been studied in the
past\cite{Naka}. No detailed information is available on the sine-Gordon theory
coupled to an additional scalar field, however, since the coupling is well in
the massive regime, it is clear that we expect bound states of these
solitons. The most likely candidate is a triplet of the same nature as in the
externally dimerized chain. If enough energy is available to overcome the
binding energy, this triplet state can then disintegrate into two solitons. We
expect then a triplet mode that is well-defined below the solitonic
continuum. This is consistent with what we observe.

To summarize, we have shown by inelastic neutron measurement that there is a
mid-gap dispersive mode and confirmed the existence of a continuum of
excitations.  We have proposed that this continuum is made of unbound S=1/2
domain walls.

It is a pleasure to thank S. Aubry, M. Azzouz, A. R. Bishop, G. J. McIntyre,
M. Poirier, H. Schulz and A. Tsvelik for interesting discussions and
M. Geoghegan for reading the manuscript.

%%%%%%%%%%%%%%%%%%%%%%%%%%%%%%%%%%%%%%%%%%%%%%%%%%%%%%%%%%%%%
$*$ Email : ain@cea.fr

%%%%%%%%%%%%%%%%%%%%%%%%%%%%%%%%%%%%%%%%%%%%%%%%%%%%%%%%%%
\narrowtext

\figure{(IN14-ILL) Three energy scans for Q=(0,1,$Q_c$) with Q$_c$=\{0.5, 0.48,
 0.46\}, at T=2.6 K. They are vertically shifted apart for clarity. The
 horizontal graduation is common to all scans; the left vertical axis is for
 Q=(0,1,0.5) scan only. Each horizontal arrow indicate the zero intensity level
 for the scan above it. In the insert a general view of the first scan
 displaying the five elements described in the text. Now labelling a dimer in
 its singlet state by $\bullet -\bullet$, the triplet state by $\Uparrow$ and a
 spin 1/2 on a copper site by $\uparrow$, we can represent the peak of the
 magnon-like mode as a traveling triplet $\bullet -\bullet\ \ \bullet -\bullet\
 \ \Uparrow\ \bullet -\bullet\ \ \bullet -\bullet$, then after the 'solitonic
 gap' the continuum would correspond to delocalized spins 1/2 such as $\bullet
 -\bullet\ \uparrow\ \bullet -\bullet\ \ \bullet -\bullet\ \uparrow\ \bullet
 -\bullet$. \label{ccont}}

\figure{(4F1-LLB) Six energy scans for Q=(0,Q$_b$,0.5) with Q$_b$=\{1, 1.2, 1.4,
1.6, 1.8, 2\} at T=1.7 K.  Same convention for axes as on
Fig~(\ref{ccont}). Maximum peak position is indicated by a vertical arrow, the
intensity reached is written below. The sharp peak of the magnon-like mode, the
'solitonic gap' and the continuum are clearly visible in all
scans.\label{bcont}}

\figure{(4F1-LLB) Three energy scans at Q=(0,1,0.5). 1- At 1.7 K (circles) we
have successively, the incoherent peak, the 'triplet gap', the magnon-like mode
that reaches 1983 counts at 2 meV, the 'solitonic gap', and the continuum. 2- At
29 K (diamonds) the continuum, similar to that of the S=1/2 HAFC. 3- At 150 K
(squares) the purely paramagnetic region. In the insert, subtraction of the scan
at 1.7 K from the one at 150 K showing the magnon-like mode and the continuum.
\label{corbose}} 
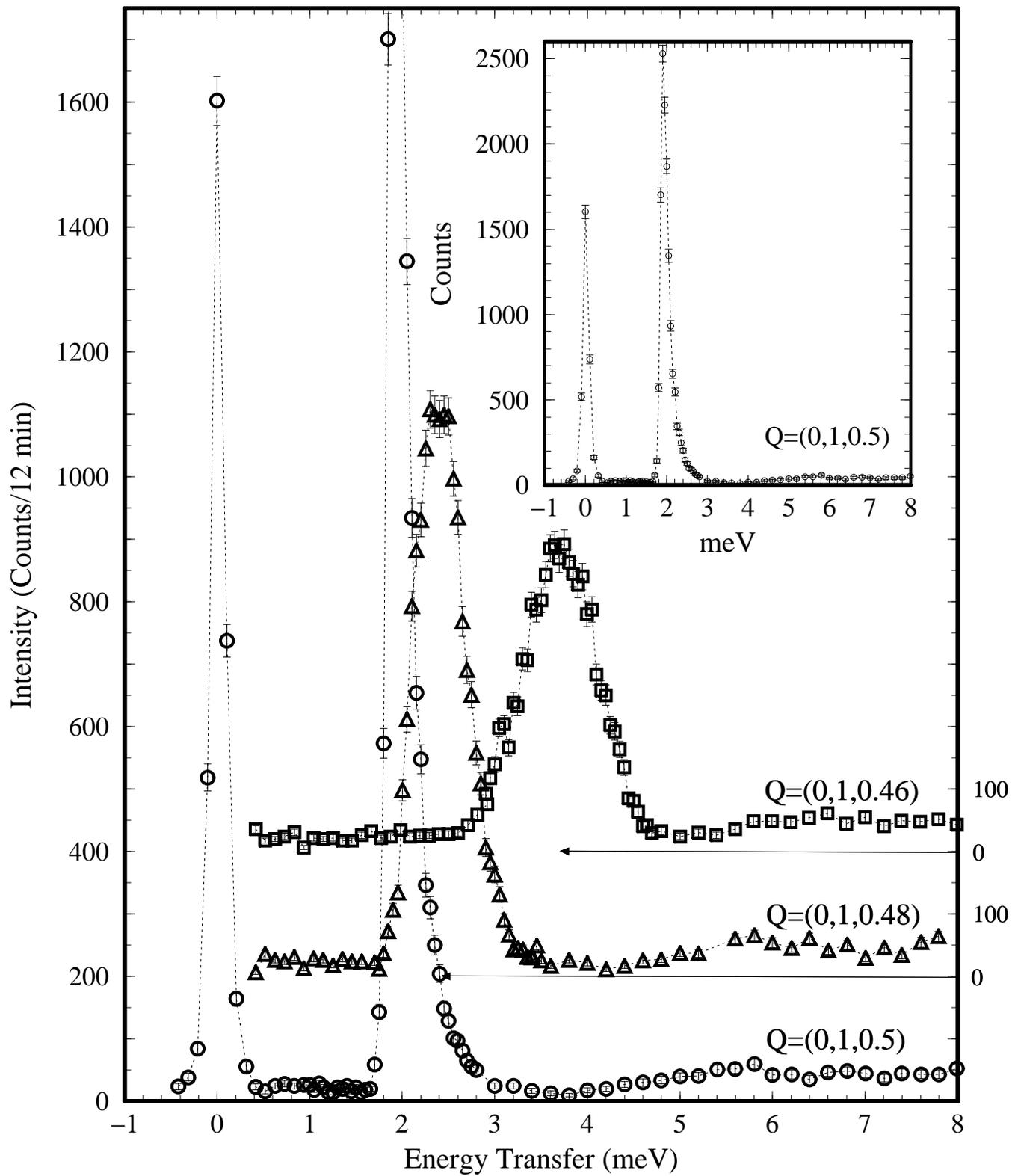
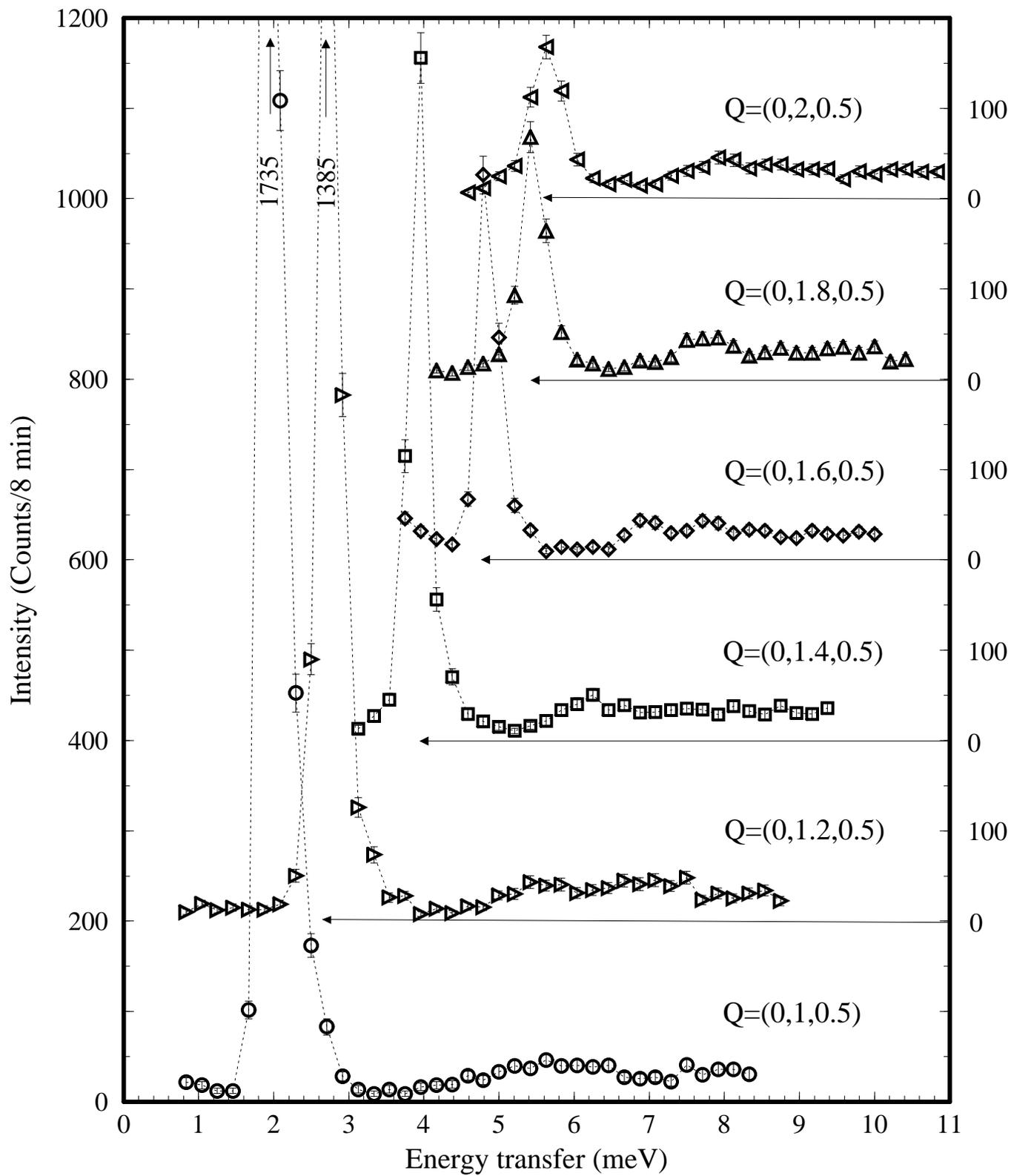
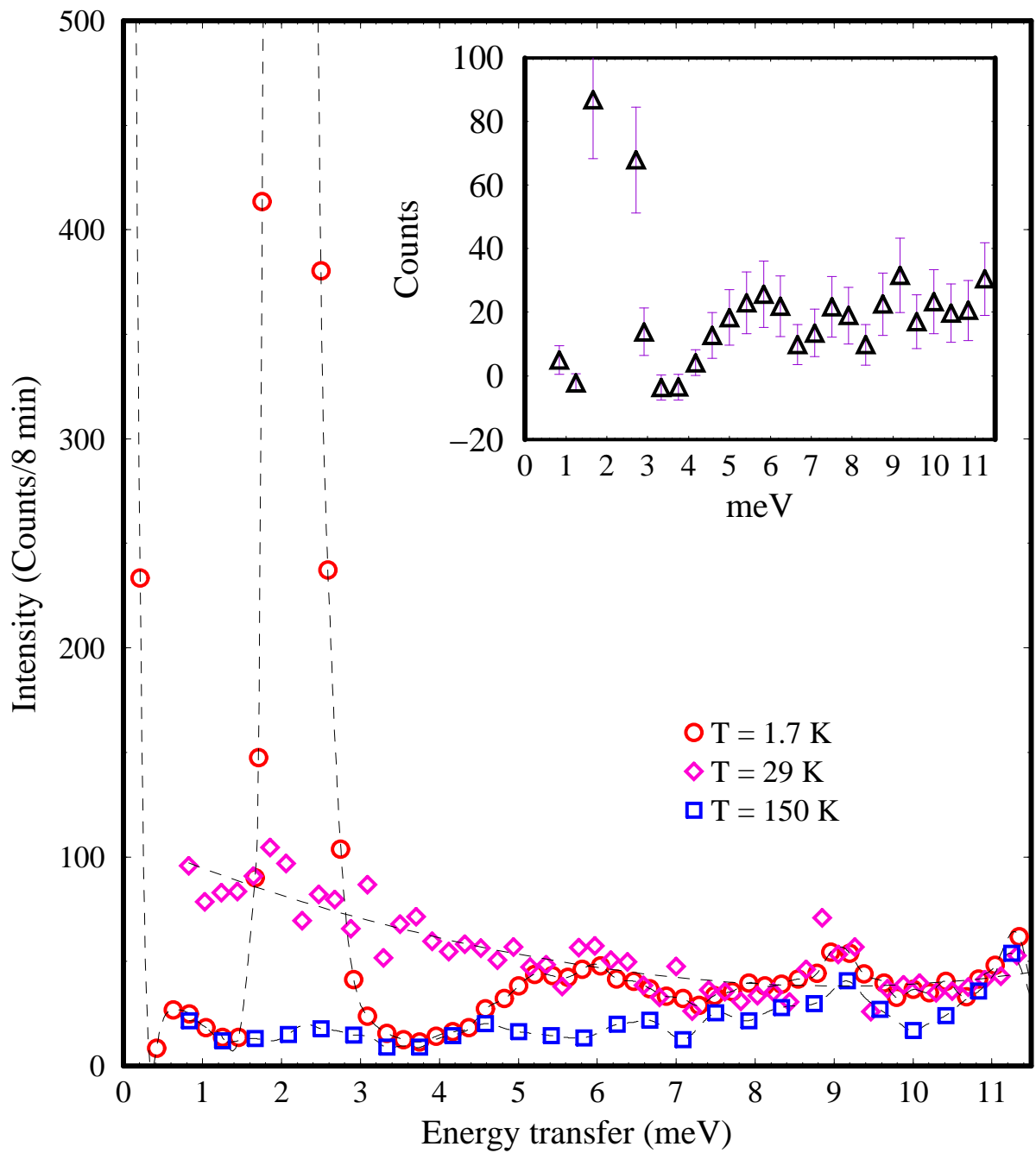
\end{document}